\documentclass{article}
\usepackage{spconf,amsmath,graphicx,hyperref}
\usepackage{booktabs}
\usepackage{multirow}
\usepackage[table]{xcolor}
\usepackage{colortbl}

\title{VoxAnon: a balanced dataset\\
for a fair evaluation of speech anonymization}
\twoauthors
  {Orane Dufour, Paul Magron}
	{Université de Lorraine, CNRS\\
	Inria, LORIA, F-54000 Nancy, France}
  {Mickael Rouvier}
	{LIA, Avignon University\\
    F-84911 Avignon, France}
    
\begin{document}
\ninept
\maketitle
\begin{abstract}
Speech anonymization is evaluated with automatic speaker verification attackers whose performance depends on the attribute distribution of both their training and evaluation data, which may bias privacy estimates for some speaker groups. To alleviate this issue, we introduce VoxAnon, a dataset built from VoxCeleb1 and~2 with training and evaluation splits balanced across gender, age, accent, and pitch. We extract and cross-check attributes using two complementary methods, balance trial lists per subgroup, and draw test/enrollment utterances from distinct video sources in order to limit linguistic and session bias. Comparing VoxAnon with unbalanced training and test data across four attackers and three anonymizers, we show that attribute imbalance changes which subgroups appear hardest to re-identify, in directions that depend on the attribute. In the balanced setting, residual disparities persist, mostly tied to accent and pitch, and depend on the anonymizer/attacker pair. Code and trial lists are publicly available.
\end{abstract}
\section{Introduction}
Speech conveys a wide range of personal information about the speaker, including cues related to speaker identity, age, gender, or health state~\cite{Kröger2020}. This raises privacy concerns that have motivated the development of speech anonymization systems, whose objective is to remove speaker-identifying information from the speech signal while preserving the information required for downstream tasks.


Speech anonymization is commonly evaluated through an automatic speaker verification (ASV) system acting as an attacker. The attacker determines whether two speech samples originate from the same speaker. In this setting, the attacker anonymizes an \textit{enrollment set} for which speaker identities are known and compares it with an anonymized \textit{test set}, assuming that some or all test speakers are represented in the enrollment set. Speaker embeddings are extracted from both sets and a pairwise comparison score, such as cosine similarity, is computed for each enrollment-test pair, which defines a \textit{trial}. Re-identification performance is typically evaluated using the equal error rate (EER), defined as the operating point where the false acceptance and false rejection rates are equal~\cite{DBLP:journals/corr/abs-2109-00648}.

Because the ASV attacker is trained on a specific dataset and may therefore inherit its biases and limitations, the resulting privacy estimates can be strongly influenced by the characteristics of both its training data and the data used to evaluate anonymization. In particular, imbalances in speaker attributes within the attacker’s training data may introduce learning biases~\cite{PERI2023101481,hutiri2022designguidelinesinclusivespeaker,10.1145/3393822.3432325}, affecting the reliability of privacy assessments. Similarly, attribute imbalances in the evaluation set may favor specific speaker groups and lead to biased and potentially unfair estimates of anonymization performance~\cite{baali-etal-2025-sveritas}.

Several studies have already investigated biases in speech anonymization. In~\cite{leschanowsky2023voiceanonymizationbias}, the authors point out that the standard Voice Privacy Challenge (VPC) test set, LibriSpeech-test~\cite{7178964}, only comes with gender labels, and addresses this limitation by curating a new benchmark built upon the guidelines of~\cite{hutiri2022designguidelinesinclusivespeaker}. However, their benchmark is restricted to only two labels, and no attention is given to the training set. In~\cite{gaznepoglu25_interspeech}, the authors conduct a per-speaker analysis and uncover linguistic biases in LibriSpeech-test. This finding is corroborated by~\cite{Franzreb2026}, who show that the vocabulary used by LibriSpeech speakers is distinctive enough to leak identity even after anonymization, and further demonstrate that spontaneous speech is comparatively more robust to this type of leakage. In~\cite{Franzreb2023ACE}, the authors evaluate three datasets that each provide different utterance and speaker-level metadata, and find that the observed disparities vary considerably across datasets. However, none of these datasets is explicitly balanced with respect to the attributes under study. Existing work thus suffers from one or both of the following limitations. First, explicit design guidelines ensuring a fair distribution of target/non-target trials across subgroups and speakers are often not followed at all, which can introduce the very bias one seeks to measure. Second, the attacker training set's balance is left unaddressed with prior work focusing almost exclusively on the construction of a test and enrollment set, leaving evaluation results similarly susceptible to bias induced by the training set.

To address these limitations, we propose VoxAnon, a dataset for speech anonymization evaluation built upon VoxCeleb1~\cite{nagrani17_interspeech} and VoxCeleb2~\cite{Vox2}. To the best of our knowledge, this is the first balanced set with respect to the marginal distributions of four speaker attributes (gender, age, accent, and pitch) on the training and test splits, and to take into account potential linguistic and video source bias. By comparing VoxAnon with attribute-imbalanced training and evaluation conditions, we show that subgroup re-identification performance varies substantially with the attribute composition of both the training and evaluation data, highlighting the importance of controlling these distributions in speech anonymization evaluation. Furthermore, using this dataset, we conduct the first subgroup evaluation of the three most recent VPC baseline anonymization systems against four attackers with different architectures. We observe disparities that are consistently tied to specific attributes and, in some cases, to specific anonymizer/attacker combinations.





\section{VoxAnon Dataset}
\label{sec:dataset}
In this section we motivate and present VoxAnon.
The code for generating the dataset, along with files containing speaker attributes and utterance lists, is available on our repository.\footnote{https://github.com/OraneD/VoxAnon}
The training and evaluation datasets' statistics are reported in Table~\ref{tab:train_and_trials_stats}.

\subsection{Speaker Attributes}
\label{sec:dataset-speaker_attributes}

Obtaining speaker attributes is necessary to construct a balanced dataset. We do this by combining two methods as follows. First, we use SpeakerCard~\cite{Peng2026speakercard} to get automatically extracted gender, age, accent, and pitch labels for all speakers in VoxCeleb1 and 2. The authors report an accuracy of 94.5\% for gender and 74.0\% for age on a 200-speaker validation subset. For accent, they report a confidence of around 80\%, and for pitch, a correlation of 0.97 across three extractors, thus these estimates appear quite reliable. Nevertheless, we then cross-check these attributes against additional sources. Specifically, we use the official VoxCeleb metadata, which provide gender and nationality for VoxCeleb1 and gender only for VoxCeleb2, as well as the web-scraped speaker attributes provided by VoxCeleb Enrichment~\cite{9688085}, which include age and nationality for speakers of VoxCeleb1 and 2. Nationality is used as a consistency check for the predicted accent labels (England = English, India = Indian, US = US, Australia/New Zealand = Australian). Similarly, when available, we verify if the age reported by VoxCeleb Enrichment falls within the corresponding predicted age range.

For VoxCeleb1, which is used to construct our test and enrollment sets (see Section~\ref{sec:dataset-test_enroll}), we adopt a conservative verification strategy. We exclude all speakers for whom an age or accent label cannot be verified, either because of a mismatch with the available metadata or because the required information is missing from VoxCeleb Enrichment. This ensures that the speaker attributes used in the evaluation are supported by at least one other source.

For VoxCeleb2, which is used for training (see Section~\ref{sec:dataset-train}) and therefore requires a substantially larger number of speakers, we apply a less restrictive criterion. We exclude speakers whose predicted accent is inconsistent with their nationality, but retain speakers for whom the age prediction cannot be verified when no age information is available in VoxCeleb Enrichment. This trade-off allows us to maintain a sufficiently large training population while still enforcing consistency for the attributes that can be reliably cross-checked.

\subsection{Attacker Training Set}
\label{sec:dataset-train}

The training set was drawn from VoxCeleb2 exclusively. We aimed for a balanced number of speakers per attribute subgroup. As described above, gender and accent labels were verified for a sufficient number of speakers to reach a balanced set of 1{,}000 speakers and around 291 hours of speech along these attributes. However, age labels proved more limiting, as too few were available in VoxCeleb Enrichment to allow cross-checking, leading us to retain the extracted age range from~\cite{Peng2026speakercard} for 554 of the 1{,}000 selected speakers. 
Note that since speakers do not all have the same number of utterances, this set exhibits a slight variability in terms of number of hours per subgroup, as observed in Table~\ref{tab:train_and_trials_stats}.

For comparison, we also generated an \textit{unbalanced} training set, for which we randomly sampled 1{,}000 VoxCeleb2 speakers for around 308 hours of speech, without controlling for speaker attributes. This results in an unbalanced dataset attribute-wise which roughly follows the speaker attribute distribution of VoxCeleb2. Details on the number of speakers, utterances and hours per subgroup for this unbalanced set are available on our repository. 

\begin{table}[t]
\centering
\scriptsize
\setlength{\tabcolsep}{2.5pt}
\renewcommand{\arraystretch}{1.08}
\caption{Number of speakers, utterances and duration in the training set, and number of test and enrollment speakers, and target (T), non-target (NT), and total trials for the evaluation set.}
\label{tab:train_and_trials_stats}
\begin{tabular}{
    l@{\hspace{4pt}}l
    >{\columncolor{gray!10}}r
    >{\columncolor{gray!10}}r
    >{\columncolor{gray!10}}r
    rrrrr
}
\toprule
& &
\multicolumn{3}{c}{\textbf{Training set}} &
\multicolumn{5}{c}{\textbf{Evaluation set}} \\
\cmidrule(lr){3-5}
\cmidrule(lr){6-10}
\textbf{Attr.} &
\textbf{Group} &
\textbf{Speak.} &
\textbf{Utt.} &
\textbf{Dur. (h)} &
\multicolumn{2}{c}{\textbf{Speak.}} &
\multicolumn{3}{c}{\textbf{Trials}} \\
 & &  & & & \textbf{Test} & \textbf{Enrol.} & \textbf{T} & \textbf{NT} & \textbf{Total}  \\
\midrule
\rowcolor{gray!20}
\textbf{All} &
\textbf{} &
\textbf{1{,}000} &
\textbf{137{,}121} &
\textbf{291.48} &
\textbf{50}&
\textbf{100}&
\textbf{500} &
\textbf{49{,}500} &
\textbf{50{,}000} \\
\addlinespace[2pt]
\multirow{2}{*}{Gender}
 & Male   & 500 & 69{,}538 & 147.23 &25 & 50& 250 & 24{,}750 & 25{,}000 \\
 & Female & 500 & 67{,}583 & 144.26 & 25& 50& 250 & 24{,}750 & 25{,}000 \\
\addlinespace[2pt]
\cmidrule(lr){1-10}
\multirow{4}{*}{Age}
 & 18--24 & 250 & 26{,}587 & 54.63 &12 &25 & 120 & 11{,}880 & 12{,}000 \\
 & 25--34 & 250 & 31{,}717 & 64.36 & 13&25 & 130 & 12{,}870 & 13{,}000 \\
 & 35--59 & 250 & 37{,}353 & 81.45 & 13&25 & 130 & 12{,}870 & 13{,}000 \\
 & 60+    & 250 & 41{,}464 & 91.05 & 12&25 & 120 & 11{,}880 & 12{,}000 \\
\addlinespace[2pt]
\cmidrule(lr){1-10}
\multirow{4}{*}{Accent}
 & US         & 263 & 39{,}132 & 85.39 &13 &26 & 130 & 12{,}870 & 13{,}000 \\
 & English    & 263 & 34{,}980 & 77.13 & 13&25 & 130 & 12{,}870 & 13{,}000 \\
 & Indian     & 263 & 36{,}899 & 72.00 &12 &25 & 120 & 11{,}880 & 12{,}000 \\
 & Austral. & 211 & 26{,}110 & 56.96 & 12&25 & 120 & 11{,}880 & 12{,}000 \\
\addlinespace[2pt]
\cmidrule(lr){1-10}
\multirow{3}{*}{Pitch}
 & Low  & 308 & 41{,}734 & 87.25  & 17&33 & 170 & 16{,}830 & 17{,}000 \\
 & Mid  & 333 & 48{,}235 & 104.67 & 17&34 & 170 & 16{,}830 & 17{,}000 \\
 & High & 359 & 47{,}152 & 99.57  & 16&33 & 160 & 15{,}840 & 16{,}000 \\
\bottomrule
\end{tabular}
\end{table}

\subsection{Evaluation Set}
\label{sec:dataset-test_enroll}

The evaluation set is constructed from VoxCeleb1 and comprises test and enrollment subsets. We first select 100 speakers to obtain a balanced attribute distribution across the test and enrollment sets, before selecting utterances to limit linguistic and video-source bias.

\subsubsection{Balancing trial counts}
\label{sec:dataset-trialcounts}
Following the guidelines of~\cite{hutiri2022designguidelinesinclusivespeaker}, a fair ASV evaluation requires the number of target and non-target trials to be equal for each speaker. Since we further aim to enable subgroup evaluation, this balance must also hold within each attribute subgroup. To this end, we design a test and enrollment set of 50 and 100 speakers respectively, drawn from VoxCeleb1, with labels evenly distributed across subgroups, as shown in Table~\ref{tab:train_and_trials_stats}. We select 10 utterances per test speaker and~5 utterances per enrollment speaker. However, the split into 3 or 4 subgroups does not always divide 50 test or 100 enrollment speakers evenly, which causes the target and non-target trial counts not to follow exactly the same ratio within every subgroup.

As for the training set (see Section~\ref{sec:dataset-train}), for comparison purposes and following the recommendation of~\cite{hutiri2022designguidelinesinclusivespeaker}, we also generated unbalanced test and enrollment sets: for each attribute, we chose a subgroup to downsample by half (male, 60+, Indian, high-pitched), and added speakers in the other groups to keep balance among them, maintaining a pool of 50 speakers in the test set and 100 speakers in the enrollment set. These specific attributes are here chosen arbitrarily: testing all possible unbalance combinations is left to future work, but our code allows to generate alternative unbalanced sets.



\subsubsection{Trial difficulty and residual bias}
\begin{figure}[t]
    \centering
    \includegraphics[width=0.85\columnwidth]{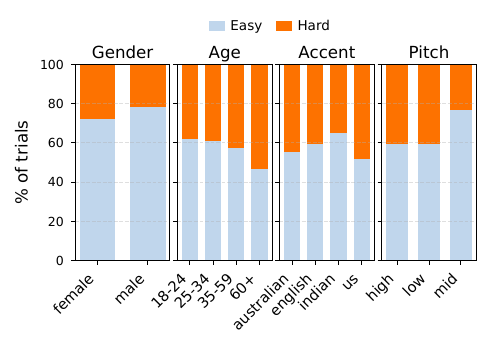}
    \vspace{-2em}
    \caption{Trial difficulty across subgroups in the balanced test set.}
    \label{fig:Combination_distribution}
\end{figure}

Although trials are balanced with respect to each \textit{individual} attribute, this does not extend to attribute \textit{combinations}: trials within a subgroup are formed by matching speakers on that single attribute only, without controlling for the other three. As a result, the proportion of  non-target trials sharing 0 to 3 additional attributes is not controlled across subgroup values, which can make some trials easier than others and introduce residual bias, as discussed in~\cite{li2022cpmapnovelevaluation}. We thus quantified this potential bias within our balanced test set by counting, for all non-target trials in each subgroup, how many of the additional attributes co-occur, grouped into \textit{easy} (0--1) and \textit{hard} (2--3), as shown in Figure~\ref{fig:Combination_distribution}. While we observe that gender and pitch skew toward easy trials, likely because these two attributes are physiologically entangled and co-occur together more often, the intra-attribute distribution remains fairly balanced overall.

\subsubsection{Limiting linguistic and source bias}
VoxCeleb speakers are recorded across multiple source videos. This property is particularly valuable for anonymization evaluation, as it lets us emulate a realistic attack scenario in which the attacker's enrollment recordings do not come from the same source as the anonymized test recordings. Selecting test and enrollment utterances from distinct video sources further prevents the ASV system from exploiting session-level context, and limits linguistic bias arising from shared vocabulary across utterances of the same video.

To enforce this as strictly as possible, we extract sentence embeddings for all trial utterances using SONAR~\cite{duquenne2023sonarsentencelevelmultimodallanguageagnostic}, a multilingual, multimodal sentence embedding model that maps text and speech into a shared representation space. For each speaker, we average the utterance embeddings belonging to the same video to obtain one embedding per video, and compute pairwise cosine similarities across all of that speaker's videos. Test and enrollment utterances are then drawn from the video pair with the lowest similarity. Since each test speaker requires 10 utterances, the least similar video pair did not always contain enough utterances. In such cases, we selected the closest video pair meeting this requirement, which occasionally led to a higher similarity. We use the same protocol to select the videos of the unbalanced set (\textit{cf}. Section~\ref{sec:dataset-trialcounts}). The resulting average video-pair cosine similarity is $0.46 \pm 0.09$ for the balanced test set and $0.45 \pm 0.09$ for the unbalanced one. For reference, the average similarity is $0.79 \pm 0.14$ across closest pairs, which indicates a clear linguistic content separation between these pair combinations.

\section{Evaluation Setup}
\label{sec:evaluation}

\begin{table*}[t]
\centering
\tiny
\caption{Results for systems trained and tested on VoxAnon-balanced. EER (\%) by attacker, anonymization condition, and speaker-attribute subgroup, and intra-attribute EER standard deviation (Std.).
}
\label{tab:eer_by_subcategory}
\resizebox{\textwidth}{!}{%
\begin{tabular}{l l r >{\columncolor{gray!12}}r >{\columncolor{gray!12}}r >{\columncolor{gray!12}}r r r r r r >{\columncolor{gray!12}}r >{\columncolor{gray!12}}r >{\columncolor{gray!12}}r >{\columncolor{gray!12}}r >{\columncolor{gray!12}}r r r r r}
\toprule
 &  & \multicolumn{1}{c}{Overall} & \multicolumn{3}{c}{Gender} & \multicolumn{5}{c}{Age} & \multicolumn{5}{c}{Accent} & \multicolumn{4}{c}{Pitch} \\
\cmidrule(lr){3-3}\cmidrule(lr){4-6}\cmidrule(lr){7-11}\cmidrule(lr){12-16}\cmidrule(lr){17-20}
Attacker & Cond. & EER & Male & Female & Std & 18--24 & 25--34 & 35--59 & 60+ & Std & US & English & Indian & Austral. & Std & Low & Mid & High & Std \\
\midrule
\multirow{4}{*}{RedimNet} & orig & \textbf{1.40} & 1.57 & 1.13 & \textbf{0.22} & 1.62 & 0.84 & 1.72 & 1.51 & \textbf{0.35} & 1.45 & 1.64 & 0.77 & 1.97 & \textbf{0.44} & 1.84 & 1.62 & 0.84 & \textbf{0.43} \\
 & B3 & \textbf{28.17} & 27.22 & 28.80 & \textbf{0.79} & 28.24 & 28.01 & 27.82 & 28.46 & \textbf{0.24} & 24.45 & 31.47 & 27.55 & 25.55 & \textbf{2.67} & 28.26 & 27.76 & 27.60 & \textbf{0.28} \\
 & B4 & \textbf{22.60} & 23.39 & 22.01 & \textbf{0.69} & 22.34 & 23.03 & 23.95 & 21.63 & \textbf{0.86} & 14.55 & 26.63 & 21.01 & 26.81 & \textbf{5.02} & 26.98 & 20.48 & 19.91 & \textbf{3.21} \\
 & B5 & \textbf{29.91} & 32.38 & 28.37 & \textbf{2.01} & 29.36 & 28.43 & 31.53 & 31.48 & \textbf{1.35} & 25.70 & 32.22 & 26.76 & 34.22 & \textbf{3.58} & 32.83 & 31.02 & 26.19 & \textbf{2.80} \\
\midrule
\multirow{4}{*}{ResNet} & orig & \textbf{1.60} & 1.36 & 1.95 & \textbf{0.29} & 1.63 & 0.85 & 1.53 & 2.51 & \textbf{0.59} & 2.27 & 2.27 & 0.85 & 0.73 & \textbf{0.74} & 1.81 & 1.22 & 1.91 & \textbf{0.30} \\
 & B3 & \textbf{30.60} & 31.10 & 30.84 & \textbf{0.13} & 32.64 & 31.55 & 29.33 & 29.35 & \textbf{1.43} & 26.03 & 33.05 & 32.44 & 28.48 & \textbf{2.89} & 31.18 & 30.60 & 30.51 & \textbf{0.30} \\
 & B4 & \textbf{26.29} & 26.46 & 25.54 & \textbf{0.46} & 25.82 & 23.99 & 27.86 & 25.60 & \textbf{1.37} & 19.67 & 29.41 & 23.28 & 30.05 & \textbf{4.32} & 30.10 & 26.44 & 22.51 & \textbf{3.10} \\
 & B5 & \textbf{31.44} & 31.97 & 30.66 & \textbf{0.65} & 30.99 & 30.59 & 32.44 & 33.16 & \textbf{1.05} & 30.66 & 31.29 & 29.98 & 34.32 & \textbf{1.66} & 32.74 & 33.04 & 29.39 & \textbf{1.65} \\
\midrule
\multirow{4}{*}{ECAPA-ssl} & orig & \textbf{3.62} & 2.10 & 4.40 & \textbf{1.15} & 2.50 & 3.17 & 2.30 & 5.84 & \textbf{1.42} & 2.26 & 5.42 & 3.42 & 1.73 & \textbf{1.42} & 2.88 & 4.13 & 3.68 & \textbf{0.52} \\
 & B3 & \textbf{28.37} & 29.52 & 27.22 & \textbf{1.15} & 31.66 & 30.91 & 24.62 & 26.48 & \textbf{2.96} & 27.07 & 30.77 & 26.76 & 25.84 & \textbf{1.88} & 30.00 & 28.32 & 26.28 & \textbf{1.52} \\
 & B4 & \textbf{18.99} & 18.47 & 19.20 & \textbf{0.37} & 18.47 & 18.46 & 14.69 & 21.67 & \textbf{2.47} & 14.62 & 23.87 & 15.82 & 19.30 & \textbf{3.59} & 21.19 & 18.89 & 16.25 & \textbf{2.02} \\
 & B5 & \textbf{29.35} & 29.20 & 29.20 & \textbf{0.00} & 33.33 & 26.93 & 29.23 & 28.33 & \textbf{2.38} & 25.39 & 34.62 & 27.50 & 29.17 & \textbf{3.42} & 31.05 & 30.42 & 26.94 & \textbf{1.81} \\
\midrule
\multirow{4}{*}{WavLM-ECAPA} & orig & \textbf{1.23} & 0.47 & 1.08 & \textbf{0.30} & 0.11 & 0.68 & 0.78 & 1.50 & \textbf{0.49} & 0.81 & 1.43 & 0.83 & 0.88 & \textbf{0.26} & 0.56 & 0.73 & 1.26 & \textbf{0.30} \\
 & B3 & \textbf{28.39} & 29.61 & 27.06 & \textbf{1.28} & 30.00 & 28.45 & 27.91 & 27.65 & \textbf{0.91} & 22.70 & 33.02 & 27.54 & 28.96 & \textbf{3.69} & 30.89 & 29.78 & 24.42 & \textbf{2.82} \\
 & B4 & \textbf{18.03} & 16.48 & 18.56 & \textbf{1.04} & 18.54 & 17.49 & 14.72 & 18.88 & \textbf{1.64} & 12.36 & 23.84 & 17.65 & 15.99 & \textbf{4.15} & 18.19 & 17.14 & 16.23 & \textbf{0.80} \\
 & B5 & \textbf{30.42} & 27.66 & 32.79 & \textbf{2.56} & 30.72 & 30.38 & 29.93 & 28.48 & \textbf{0.85} & 27.90 & 28.73 & 29.02 & 30.72 & \textbf{1.03} & 27.75 & 31.65 & 31.33 & \textbf{1.77} \\
\bottomrule
\end{tabular}%
}
\end{table*}

\subsection{Attack scenario and systems}

To assess the re-identification risk in a worst-case scenario, we consider semi-informed~\cite{informedattackers} attackers, i.e., they are trained on data anonymized in the same manner as the test and enrollment data. The only difference is the source-to-target speaker mapping used during the voice conversion stage of the anonymizer, which the attacker cannot realistically have access to, as this mapping is random. Motivated by prior work~\cite{dufour2026largescaleperspeakeranalysisreidentification} that showed that per-speaker risk considerably varies as a function of the anonymizer and the attacker, we consider 3 anonymizers, plus original speech, and 4 different attackers, which yields a total of 16 attack scenarios. Both the balanced and unbalanced VoxAnon-train sets are first anonymized using each possible anonymizer, in order to further train the attackers.

We consider the following attackers. \textbf{ResNet}~\cite{resnet, kiwano2026} is a strong ASV baseline that extracts speaker embeddings based on residual blocks (we use the deeper ResNet-101 version). \textbf{RedimNet}~\cite{redimnet} reshapes 2D spectrogram feature maps into 1D representations and back, enabling joint 1D/2D residual blocks. \textbf{WavLM-ECAPA} combines self-supervised WavLM~\cite{wavlm} representations with an ECAPA-TDNN~\cite{desplanques20_interspeech} speaker embedding head. Lastly, \textbf{ECAPA-ssl}~\cite{ecapa_ssl} is the baseline attacker of the latest VPC~\cite{eurecomVPC206}. It fuses WavLM and filter-bank features before attentive pooling into an ECAPA-TDNN, fine-tuned on anonymized data.

For the anonymizers, we use three standard baselines from the latest VPC~\cite{eurecomVPC206}.
\textbf{B3}~\cite{B3_2023} extracts phonetic transcriptions, F0, energy, and phone durations. It then perturbs F0 and energy per phone, to finally resynthesize speech with and artificially generated speaker embedding passed through and HiFi-GAN vocoder. \textbf{B4}~\cite{B4} is a neural audio codec language modeling system that renders semantic and acoustic tokens, and replaces speaker identity with a pseudo-speaker voice drawn from a predefined pool. \textbf{B5}~\cite{B5} extracts vector quantized bottleneck features from a Wav2Vec 2.0 model, combined with F0 and a target-speaker one-hot vector, fed to a HiFi-GAN vocoder which replaces the voice of the input speaker by the voice of a speaker drawn from its training set.

\subsection{Bias and fairness evaluation}
Group fairness refers to the absence of discrimination associated with a demographic group defined by a protected attribute, such as gender or age~\cite{benítezpeña2026tradeoffsindividualgroupfairness}. Such a discrimination is assessed by computing metrics across demographic subgroups to reveal performance disparities. While there exists an extensive literature on ASV fairness metrics~\cite{fenu21_interspeech, chouchane24_odyssey}, establishing and studying the behavior of such metrics is not the focus of this work.
Therefore, to keep things simple in this paper, we evaluate fairness by computing the standard deviation of the EER across subgroups for each attribute.

In order to assess the impact of training and evaluation set imbalance on subgroup bias, we focus on the undersampled subgroup within each attribute of our unbalanced test set (these are: male, 60+, Indian, high pitch). We check whether this subgroup obtains the highest (\textit{hardest} to re-identify) and lowest (\textit{easiest} to re-identify) EER among all subgroups within its category, and report the percentage of attack scenarios for which this holds.
We compare these values to the percentage obtained on the balanced test set, and we conduct this analysis for each training dataset.

\section{Results}
\label{sec:results}

\subsection{Balanced and unbalanced sets comparison}

\begin{figure*}[t]
    \centering
    \vspace{-1em}
    \includegraphics[width=\textwidth, height=4.5cm]{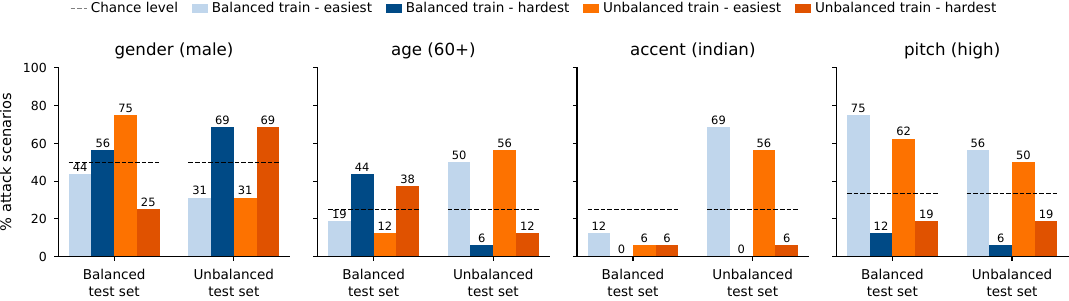}
    \vspace{-2em}
    \caption{Balanced and unbalanced sets comparisons: percentage of attack scenarios for which the undersampled subgroup for each category obtained the highest and lowest EER, corresponding to the hardest and easiest subgroup to re-identify, respectively.
    }
    \label{fig:comparison_hardest}
    \vspace{-1em}
\end{figure*}

We first investigate how dataset imbalances impact re-identification. Results are shown in Figure~\ref{fig:comparison_hardest}.
Gender shows the expected pattern: with unbalanced training, the majority class (male), is the easiest to re-identify on the balanced test (75\%), suggesting the attacker indeed models these speakers better. But when undersampled in the test set, male speakers become harder to re-identify (69\%): unbalanced training does not compensate for a skewed test set. Balanced training removes the bias on the balanced test (44/56\% $\sim$6 points from chance), but leaves the unbalanced test bias unchanged, confirming that this compensation only holds when the test itself is balanced. Accent behaves similarly: there is almost no bias on the balanced test set (0-12\%) regardless of the training conditions, but a strong bias on the unbalanced test set favoring easier re-identification of Indian-accented speakers. Balanced training actually intensifies this bias ($56\rightarrow69\%$), since Indian speakers are more represented in this set than in the unbalanced training condition. Age and pitch depart from this pattern. On the balanced test, balanced training conditions worsen the 60+ bias, rather than mitigating it, making these speakers more often harder to re-identify ($38\%\rightarrow44\%$), while on the unbalanced test, it barely reduces the bias ($56\%\rightarrow50\%$). Similarly, pitch shows a persistent bias toward easier re-identification of high-pitched speakers, both on the unbalanced and balanced test set, a bias that balanced training widens instead of reducing.

Overall, these results show that subgroup bias in anonymization evaluation can stem from either the training set, the test set, or both, depending on the attribute considered, and that balanced training does not always mitigate bias, as seen for age and pitch. This confirms that both sets must be balanced to guarantee a fair privacy evaluation. These attribute-specific effects open the way to further experiments to understand why some attributes resist this compensation, an investigation that VoxAnon is precisely designed to enable.

\subsection{Analysis under the balanced setting}
Having established that training/test set imbalance can introduce subgroup bias, we now turn to the fully balanced setting (train and test sets) to isolate disparities that stem purely from the anonymizer and attacker, independently of any dataset-level imbalance.

Table~\ref{tab:eer_by_subcategory} reports the resulting subgroup EERs. Results show that the anonymizer ranking is stable across attackers, i.e., B5 $>$ B3 $>$ B4 with average EERs of 30.28\%, 28.88\%, and 21.48\%, respectively. B3 and B5 vary little across attackers (spread of~2\%), but B4 diverges markedly: SSL-based attackers (i.e., ECAPA-ssl and WavLM-ECAPA) score roughly 8 points below RedimNet and ResNet, indicating a specific vulnerability of B4 to SSL-based attacks rather than a uniform weakness.
Subgroup dispersion is small for gender and age, with an Std. below 1.5 with isolated peaks at 2.56 and 2.96. However, it is markedly higher for accent, peaking at 5.02 for RedimNet and B4. B4 has the highest accent dispersion for every attacker, coinciding with its weaker average protection.
Among accents, US speakers are the easiest to re-identify in 10 of 12 attacker/anonymizer combinations, making accent the attribute with the most stable intra-attribute ordering. Age shows no comparably stable pattern, e.g., the 18--24 group is hardest in only 5 of 12 combinations. Gender disparity also varies with attacker and system: females are easier to re-identify than males in most combinations, but this reverses for WavLM-ECAPA.

These results demonstrate that even in an attribute-balanced setting, subgroup disparities persist in terms of re-identification risks. Accent stands out as the most consistently affected attribute. Gender and age disparities are smaller in comparison, but more dependent on the attacker and anonymizer systems.

\section{Conclusion}
\label{sec:conclusion}

We introduced VoxAnon, a training and evaluation dataset for speech anonymization balanced across speaker attributes. Comparing attacks on VoxAnon with those on an unbalanced dataset, we showed that speaker-attribute imbalances indeed bias privacy evaluation. Across several attack scenarios, we further found that residual attribute-related disparities persist despite balanced data, with their extent depending on both the anonymizer and attacker. Beyond enabling fairer anonymization systems, VoxAnon supports attribute inference research, which currently lacks labeled, balanced benchmarks. As future work, we will train attribute extractors on VoxAnon to study their benefit to attackers.

\section{Acknowledgements}
This work was supported by the French Agence Nationale de la Recherche via the SpeechPrivacy project (ANR-23-CE23-0022). It was provided with computer and storage resources by GENCI at IDRIS thanks to the grant 2025-AD011015838R1 on the supercomputer Jean Zay's V100 partition. The authors would also like to thank the late Emmanuel Vincent, whose inspiration laid the foundations of this work.

\bibliographystyle{IEEEtrans}
\bibliography{refs}

\end{document}